\documentclass[aps,prl,twocolumn]{revtex4-1}

\begin{document}
\title{Comment on "Late Time Behavior of false Vacuum Decay: \\Possible Implications for Cosmology and Metastable Inflating States"}
\author{K. Urbanowski}
\email[]{e--mail: K.Urbanowski@proton.if.uz.zgora.pl}
\affiliation{
Institute of Physics,
University of Zielona Gora,  \\
ul. Prof. Z. Szafrana 4a,
65-516 Zielona Gora,
Poland}
\pacs{98.80.Cq, 11.10.St, 98.80.Qc, 11.10.-z}
\maketitle

In the Letter \cite{Krauss} Krauss and Dent analyze late time behavior of false vacuum decay
and discuss its possible cosmological implications. Their attention is focused on the possible
behavior of the unstable false vacuum at very late times, where deviations from the exponential
decay law become to be dominat.  In my opinion the discussion presented in the Letter
requires  some complements.

Let us start from a brief introduction  into the problem. If $|M\rangle$ is an initial unstable
state then the survival probability, ${\cal P}(t)$, equals ${\cal P}(t) = |a(t)|^{2}$, where
$a(t)$ is the survival amplitude, $a(t) = \langle M|M;t\rangle$, and $|M;t\rangle =
e^{\textstyle{-itH}}\,|M\rangle$, $H$ is the total Hamiltonian of the system under considerations.
The spectrum, $\sigma(H)$, of $H$ is assumed to be bounded from below, $\sigma(H) =[E_{min},\infty)$
and $E_{min} > -\infty$. Searching for late time properties of unstable states  one usually uses the
integral representation of $a(t)$ as the Fourier transform of the energy distribution function,
$\rho(E)$, (see (2) in \cite{Krauss}), with $\rho (E) \geq 0$ and $\rho (E) = 0$ for $E < E_{min}$.
In the case of quasi--stationary (metastable) states it is convenient to express $a(t)$ in the
following form \cite{Sluis,muga-1,urbanowski-2-2009,urbanowski-1-2009}, $a(t) = a_{exp}(t) + a_{non}(t)$,
where $a_{exp}(t)$ is the exponential part of $a(t)$, that is $a_{exp}(t) =
N\,\exp\,[{-it(E_{M} - \frac{i}{2}\,\Gamma_{M})}]$,
($E_{M}$ is the energy of the system in the state $|M\rangle$ measured at the canonical decay times,
$\Gamma_{M}$ is the decay width, $N$ is the normalization constant), and $a_{non}(t)$ is the
non--exponential part of $a(t)$. From the literature it is known that $a_{non}(t)$ exhibits inverse
power--law behavior at the late time region.  The crossover time $T$
can be found by solving the following equation, $|a_{exp}(t)|^{\,2} = |a_{non}(t)|^{\,2}$.

The authors of the Letter find the crossover time $T$ (see (7) in \cite{Krauss}) and then they find
a quantitative estimation of $T$, (see (11) in \cite{Krauss}). It should be noticed that this estimation
can not be considered as conclusive because the formulae for $T$
depend on the model considered (i.e. on $\rho(E)$) in general and they can differ
from  relation (7) (see, eg.  \cite{Sluis,Sluis,muga-1,urbanowski-2-2009,urbanowski-1-2009}).
In order to find a proper relation for $T$ one needs a more realistic expression for $\rho(E)$
corresponding to the unstable false vacuum state. One can not exclude that $T$ for more realistic
$\rho(E)$ may be much shorter than that found in \cite{Krauss}, which could change the conclusions
contained there. So,  result (11) in \cite{Krauss} should be considered only as a very  rough approximation.

Another, more important remark concerns the energy of the false vacuum state  at late times $t \gg T$.
In \cite{Krauss} it is hypothesized that some false vacuum regions do survive well up to the time $T$ or  later.
Let $E^{false}_{0}$ be the energy of a state corresponding to the false vacuum measured at the canonical decay time
and $E^{true}_{0}$ be the energy of true vacuum (i.e. the true ground state of the system).
The problem is that the energy of those false vacuum regions which survived up to $T$ and much later
differs from $E^{false}_{0}$. It follows from properties of the instantaneous energy ${\cal E}_{M}(t)$ of
a unstable state $|M\rangle$, \cite{urbanowski-2-2009,urbanowski-1-2009}. In the considered case,
${\cal E}_{M}(t)$ can be found using the effective Hamiltonian,  $h_{M}(t)$, governing the time evolution in
a subspace of states spanned  by vector $|M\rangle$, \cite{urbanowski-2-2009,urbanowski-1-2009}:
\begin{eqnarray}
h_{M}(t) = \frac{i}{a(t)}\,\frac{\partial a(t)}{\partial t}
\equiv  \frac{\langle M|H|M;t\rangle}{\langle M|M;t\rangle}. \label{h2}
\end{eqnarray}
The instantaneous energy ${\cal E}_{M}(t)$ of the system in the state $|M\rangle$  is the real part of
$h_{M}(t)$, ${\cal E}_{M}(t) = \Re \,(h_{M}(t))$. There is ${\cal E}_{M}(t)= E_{M}$ at the canonical decay
time and,
\begin{equation}
{\cal E}_{M}(t) \simeq E_{min} + \frac{c_{1}}{t} \,+\,\frac{c_{2}}{t^{2}} \ldots, \;\;\;({\rm for}
\;\;t \gg T), \label{E(t)}
\end{equation}
where $c_{i} = c_{i}^{\ast},\;i=1,2,\ldots$, and $\lim_{t \rightarrow \infty}\, {\cal E}_{M}(t) = E_{min}$.
So, if  one assumes that $E^{true}_{0} \equiv E_{min}$ then one has for the false vacuum state that at $t \gg T$
\begin{equation}
E^{false}_{0}(t) \simeq E^{true}_{0} + \frac{c_{1}}{t} + \frac{c_{2}}{t^{2}}\ldots \;\; .
\label{E-false-infty}
\end{equation}

The basic physical factor forcing the wave function  $|M;t\rangle$ and thus the  amplitude $a(t)$
to exhibit inverse power law behavior at $t \gg T$ is a boundedness from below of  $\sigma (H)$. This means
that if this condition takes place and $\int _{-\infty}^{+\infty}\rho(E)\,dE\,<\,\infty$, then all  properties
of $a(t)$, including a form of the time--dependence at  $t \gg T$, are the  mathematical consequence of them both.
The same applies by (\ref{h2}) to properties of $h_{M}(t)$ and concerns
the asymptotic form of $h_{M}(t)$ and
thus of ${\cal E}_{M}(t)$ at $t \gg T$. Note that properties of $a(t)$ and $h_{M}(t)$ discussed above
do not take place when  $\sigma(H) = (-\infty, + \infty)$. In conclusion,
above considerations show that the results concerning the late time behavior of the false vacuum regions
obtained in \cite{Krauss} should be modified using properties (\ref{E(t)}), (\ref{E-false-infty}).

\end{document}